\documentclass[prd,preprint,preprintnumbers,amsmath,amssymb]{revtex4}

\usepackage{bm,curves}
\usepackage{epsfig}
\usepackage[colorlinks = true, linkcolor = blue]{hyperref}

\begin{document}

\preprint{JLAB-THY-12-1645}

\title{Parton distributions in the presence of target mass corrections}

\author{F. M. Steffens$^1$, M. D. Brown$^2$, W. Melnitchouk$^3$,
	S. Sanches$^1$}
\affiliation{
	$^1$\mbox{IF -- USP, Cidade Universitaria S\~ao Paulo, SP, Brazil}\\
	$^2$\mbox{Department of Physics, Arizona State University,
		Tempe, Arizona 85287, USA} \\
	$^3$\mbox{Jefferson Lab, 12000 Jefferson Avenue,
		Newport News, Virginia 23606, USA}}

\date{\today}

\begin{abstract}
We study the consistency of parton distribution functions in the
presence of target mass corrections (TMCs) at low $Q^2$.  We review
the standard operator product expansion derivation of TMCs in both
$x$- and moment-space, and present the results in closed form for
all unpolarized structure functions and their moments.  To avoid
the unphysical region at $x > 1$ in the standard TMC analysis,
we propose an expansion of the target mass corrected structure
functions order by order in $M^2/Q^2$, and assess the convergence
properties of the resulting forms numerically.
\end{abstract}

\maketitle

%%%%%%%%%%%%%%%%%%%%%%%%%%%%%%%%%%%%%%%%%%%%%%%%%%%%%%%%%%%%%%%%%%%%%%%%%
\section{Introduction}

The application of the operator product expansion (OPE) to the 
phenomenological study of Quantum Chromodynamics (QCD) has been very
successful in the determination of the quark and gluon substructure of
the nucleon.  The OPE allows the formal separation of cross sections
for high-energy processes such as deep-inelastic scattering (DIS) into
perturbatively calculable partonic cross sections and nonperturbative
contributions parametrized by parton distribution functions (PDFs).
The factorization of the cross section becomes especially clean in the
Bjorken limit, where the energy $\nu$ and four-momentum squared $Q^2$
transferred to a nucleon with mass $M$ both become infinite, with the
ratio $x = Q^2/2M\nu$ fixed.

In the Bjorken limit the DIS process becomes dominated by scattering
at light-cone space-time distances $z_\mu z^\mu \sim 0$, with the
expansion made in terms of products of singular and non-singular
terms around the light-cone.  The singularities are isolated in the
perturbative Wilson coefficients, while the non-singular terms are all
the possible operators allowed by the underlying quantum field theory.
The coefficient of the operators of lowest twist (where twist is defined
as the dimension minus the spin of the operator) contains the most
singular terms.  Operators in the expansion with higher twist are less
singular, and at large $Q^2$ are suppressed by powers of $1/Q^2$.

While this framework has met with considerable success in describing
data at high $Q^2 \gg M^2$ and large final state hadron masses
$W^2 = M^2 + Q^2 (1-x)/x$, many recent high-precision experiments
\cite{Christy11} have been performed at lower energies, with $Q^2$
down to $\approx 1-2$~GeV$^2$, where the use of the asymptotic
Bjorken limit formalism is more questionable.  In addition to the
strong coupling constant $\alpha_s$ becoming large, at low $Q^2$
the higher twist power corrections, which describe nonperturbative
multi-parton correlations, become increasingly important.
Furthermore, even at leading twist, there are corrections arising
from purely kinematic effects associated with finite values of
$Q^2/\nu^2 = 4 M^2 x^2/Q^2$, usually termed target mass corrections
(TMCs) \cite{GP76, DGP77, DGPannals, Sch08, Bra11}.

To perform reliable perturbative QCD based analyses which include
data in the low $Q^2$ region, a careful treatment of the subleading
$1/Q^2$ corrections is essential, and global PDF analyses
\cite{ABKM, CJ10, CJ11, ABM12, NNPDF} have only recently begun
to take such effects systematically into account.
Studies of quark-hadron duality \cite{BG70, MEK} have also strongly
suggested that data at low $W$ can be described (to within
$\sim 10$--15\%) by leading twist parton distributions.
A more basic question, however, is whether one can consistently
define leading twist parton distributions in the presence of TMCs,
that can be valid at low $Q^2$ over the entire range of~$x$.

The first analysis to tackle this question was by Georgi and Politzer
(GP) \cite{GP76}, who proposed taking TMCs into account by defining
distributions at low $Q^2$ in terms of the Nachtmann scaling variable,
$\xi$ \cite{Nac73, Gre71},
\begin{eqnarray}
\xi &=& \frac{2 x}{1 + \rho},\ \ \
\textrm{with}\ \ \rho = \sqrt{1 + 4 \mu x^2}\ \ \
\textrm{and}\ \ \mu = {M^2 \over Q^2}.
\label{eq:xi}
\end{eqnarray}
This leads to a specific prescription for removing TMCs from measured
structure functions that has been used extensively in the literature
\cite{Sch08}.

Unfortunately, problems with the standard TMC prescription were soon
realized \cite{Ell76, Bar76, GTW77, BJT79, JT80, Fra80} in the behavior
of the target mass corrected structure functions in the vicinity of
$x \approx 1$.  In particular, functions expressed in terms of $\xi$
over the interval $0 \leq \xi \leq 1$ necessarily extend into the
unphysical region between the elastic limit $\xi = \xi_0 \equiv \xi(x=1)$
and $\xi = 1$ for any finite value of $Q^2$ \cite{GP76}.
This not only violates the conservation of energy and momentum,
but also makes structure functions nonzero at $x=1$, at odds with
the expectation that leading twist functions should vanish at the
elastic point \cite{BG70, Mel01}.

As a possible remedy, De~Rujula {\em et al.} \cite{DGP77, DGPannals}
noted that in the threshold region analyses of data should not be
performed in terms of leading twist structure functions alone,
without also incorporating the effects of higher twist operators.
They argued that a nonuniformity in the limits as $n \to \infty$
and $Q^2 \to \infty$ renders the entire approach untenable at very
low $W$, when higher twists exceed $\sim n M^2/Q^2$ for the $n$-th
structure function moment.

Attempts were also made by Tung and collaborators \cite{BJT79, JT80}
to phenomenologically remove the threshold problem by utilizing an
{\em ansatz} to smoothly merge the moments in the perturbative
region at large $Q^2$ with their correct threshold behavior in the
$n \to \infty$ limit, although such a prescription is not unique.
Steffens and Melnitchouk \cite{SM06} extended this approach by
proposing threshold-dependent distributions which exactly satisfy
threshold kinematics at all $Q^2$, at the expense of sacrificing
the universality of PDFs in the presence of TMCs.

Other approaches based on collinear factorization, starting with the
seminal work of Ellis, Furmanski and Petronzio \cite{EFP83}, avoid the
inversion of moments by implementing TMCs directly in momentum space
within the parton model \cite{AOT94, KR02, KR04, AQ08}.  These, too,
however, suffer from prescription dependence \cite{AOT94, KR02, KR04,
AQ08}, or do not extend to all orders in $1/Q^2$ \cite{EFP83}.
In addition, even though they do not invoke distributions at $x > 1$,
all these formulations nevertheless retain the problem of nonvanishing
structure functions at $x = 1$.

Given the desire to maximally utilize the recent precision structure
function measurements at large $x$ \cite{Christy11, ABKM, CJ10, CJ11,
ABM12, NNPDF, BONUS}, as well as those planned for the near future
\cite{BONUS12, MARATHON}, there is a pressing need to address the
question of TMCs and the consistency of parton distributions with mass
corrections at finite $Q^2$.  A more reliable treatment of the high-$x$
region at moderate $Q^2$ is important not only in providing a better
understanding of the quark structure of the nucleon in the deep valence
region \cite{MT96, Hol10}, it is also vital for constraining cross
sections at collider energies through the evolution to lower $x$ at
higher $Q^2$ values \cite{Kuh00, Bra12}.

In this paper we revisit the problem of kinematic thresholds and PDF
definitions in the OPE approach to TMCs, elucidating its shortcomings,
and proposing an alternative method that addresses some of the problems
inherent in the standard TMC formulation.
In Sec.~\ref{sec:ope} we review the standard TMC approach, outlining
the OPE derivation of target mass corrected moments and their inversion
to $x$-space.  We demonstrate explicitly the conflict of the usual
inversion procedure with energy-momentum conservation, and illustrate
its consequences for the $x$ dependence of the structure functions as
well as their moments.
We propose a new method to compute TMCs in Sec.~\ref{sec:series}, based
on inversion of the moments order by order in $M^2/Q^2$, without having
to introduce the Nachtmann scaling variable $\xi$, and study the
convergence of the series numerically.  The advantages and limitations
of this method are summarized in Sec.~\ref{sec:conc}.
Further technical details of the TMC derivations of moments and
structure functions are provided in Appendices~\ref{app:mom} and
\ref{app:sf}, respectively.

%%%%%%%%%%%%%%%%%%%%%%%%%%%%%%%%%%%%%%%%%%%%%%%%%%%%%%%%%%%%%%%%%%%%%%%%%
\section{Target mass corrections in the OPE}
\label{sec:ope}

In this section we begin by summarizing the basic formulas for
inclusive cross sections and structure functions, before outlining the
main steps in the derivation of TMCs from the operator product expansion.
We present results for the complete set of leading twist moments of
unpolarized structure functions, and discuss their inversion to obtain
the $x$ dependence at nonzero $M^2/Q^2$.

In the one-boson exchange approximation, the differential cross section
for a lepton scattering from a nucleon target is given (in the target
rest frame) by
\begin{eqnarray}
\frac{d^2 \sigma}{d\Omega \hspace{2 pt} dE'}
&=& \frac{\alpha^2}{Q^4} \frac{E'}{ME}\,
    \eta\, L_{\mu \nu} W^{\mu \nu},
\end{eqnarray}
where $\Omega$ is the scattered lepton solid angle, and $E$ and $E'$
are the initial and final electron energies, respectively.
The lepton tensors $L_{\mu \nu}$ and the coefficients $\eta$ depend
on the type of boson exchanged ($\gamma, \gamma Z, Z$, or $W^\pm$)
\cite{PDG}.
Denoting the initial and final lepton momenta by $k$ and $k'$, 
respectively, and the momentum transferred to the nucleon by 
$q = k - k'$, the hadronic tensor is given by the commutator of
electroweak current operators $J_\mu$,
\begin{align}
W^{\mu \nu}
&= \frac{1}{2\pi} \int d^4z\, e^{i q \cdot z}
   \langle N | [J^\mu(z), J^\nu(0)] | N \rangle		\\
&= -g^{\mu \nu} F_1 + \frac{p^\mu p^\nu}{p \cdot q} F_2
 - i \epsilon^{\mu \nu \lambda \sigma}
   \frac{p_\lambda p_\sigma} {2 p \cdot q} F_3
 + \frac{2q^\mu q^\nu}{Q^2} F_4
 + \frac{p^\mu q^\nu + p^\nu q^\mu}{p \cdot q} F_5,
\end{align}
where $F_i$ ($i=1-5$) are the structure functions of the nucleon,
usually expressed in terms of the variables $x$ and $Q^2 = -q^2$,
and we adopt the convention $\epsilon_{0123} = 1$ \cite{PDG}.
The structure functions $F_1$ and $F_2$ are accessible in charged lepton
or neutrino scattering through a product of vector currents, while $F_3$
requires the interference of vector and axial vector currents.
The vector structure functions $F_4$ and $F_5$ are accessible in
neutrino scattering but are suppressed by lepton masses, $m_l^2/M^2$;
for completeness, however, we include them in this analysis.
The hadronic tensor $W_{\mu \nu}$ is also related to the imaginary
part of the virtual forward Compton scattering amplitude,
\begin{equation}
T^{\mu \nu} = i \int d^4z\, e^{i q \cdot z}\,
	      \langle N | T(J^\mu(z) J^\nu(0)) | N \rangle,
\end{equation}
with 
\begin{equation}
W^{\mu \nu} = \frac{1}{\pi}\, \mathrm{disc} \hspace{2 pt} T^{\mu \nu}.
\label{eq:disc}
\end{equation}
In the following we derive expressions for the amplitude $T^{\mu \nu}$
and use Eq.~(\ref{eq:disc}) to extract the results for the structure
functions.

% .......................................................................
\subsection{Moments of structure functions}
\label{ssec:moments}

The standard derivation of TMCs in the OPE in the twist-2 approximation
begins with the Compton scattering amplitude $T_{\mu \nu}$, which can
in general be written as \cite{GP76}
\begin{eqnarray}
T^{\mu \nu}
&=& \sum_{k = 1}^\infty
  \Big(
    - g^{\mu \nu} q_{\mu_1} q_{\mu_2} C^{2k}_1
    + g^\mu_{\mu_1} g^{\nu}_{\mu_2} Q^2 C^{2k}_2
    - i \epsilon^{\mu \nu \alpha \beta}
	g_{\alpha \mu_1} q_{\beta} q_{\mu_2} C^{2k}_3
    + \frac{q^\mu q^\nu}{Q^2} q_{\mu_1} q_{\mu_2} C^{2k}_4
							\nonumber\\
& & \hspace*{1cm}
    +\ \left( g^\mu_{\mu_1} q^\nu q_{\mu_2}
	   + g^\nu_{\mu_1} q^\mu q_{\mu_2}
      \right) C^{2k}_5
  \Big)\,
  q_{\mu_3} \cdots q_{\mu_{2k}}\, \frac{2^{2k}}{Q^{4k}}\,
  A_{2k}\, \Pi^{\mu_1 \cdots \mu_{2k}},
\label{eq:T}
\end{eqnarray}
where
\begin{eqnarray} \label{eq:Pi}
\Pi^{\mu_1 \cdots \mu_{2k}}
&=& \sum_{j = 0}^k (-1)^j \frac{(2k - j)!}{2^j (2k)!}\,
  \{g \cdots g \hspace{1 pt} p \cdots p \}_{k, j}\, (p^2)^j
\end{eqnarray}
and $\{g \cdots g \hspace{1 pt} p \cdots p \}_{k, j}$ represents the
(symmetric) sum of $(2k)!/[2^j j! (2k - 2j)!]$ distinct products
of the form
$g^{\mu_{i_1} \mu_{i_2}} \cdots g^{\mu_{i_{2j -1}} \mu_{i_{2j}}}
 p^{\mu_{2j + 1}} \cdots p^{\mu_{2k}}$
resulting from permutations of the indices $\mu_1, \cdots, \mu_{2k}$.
The Wilson coefficients $C^{2k}_i$ are calculated perturbatively,
while the factors $A_{2k}$ are matrix elements of local twist-2
operators ${\cal O}^{\mu_1 \cdots \mu_{2k}}$ \cite{Bur80},
\begin{eqnarray}
\langle N |\, {\cal O}^{\mu_1 \cdots \mu_{2k}}\, | N \rangle
&=& A_{2k}\, p^{\mu_1} \cdots p^{\mu_{2k}}\ -\ {\rm traces},
\label{eq:A2k_def}
\end{eqnarray}
which parametrize the nonperturbative structure of the nucleon.
In the case of the flavor singlet operator, for example, one has
\begin{eqnarray}
{\cal O}^{\mu_1 \cdots \mu_{2k}}_{\rm sing}
&=& \bar\psi \gamma^{ \{ \mu_1 } D^{\mu_2} \cdots D^{\mu_{2k} \}} \psi\
 -\ {\rm traces},
\end{eqnarray} 
where the braces $\{ \cdots \}$ denote symmetrization with respect
to the indices $\mu_1, \cdots, \mu_{2k}$.

The Cornwall-Norton moments $M^{(n)}_i$ of the structure functions
$F_i$ are defined by 
\begin{equation}
  M^{(n)}_i(Q^2)
= \left\{
  \begin{array}{ll}
    {\displaystyle\int_0^1} dx\, x^{n - 1} F_i(x,Q^2)
	& \hspace{5 pt} \mbox{if}\ i = 1, 3, 4, 5	\\
	&						\\
    {\displaystyle\int_0^1} dx\, x^{n - 2} F_i(x,Q^2)
	& \hspace{5 pt} \mbox{if}\ i = 2, L,
  \end{array}
  \right.
\label{eq:momentdef}
\end{equation}
where the longitudinal structure function $F_L$ is given by 
\begin{equation}
F_L = (1 + 4 \mu x^2) F_2 - 2x F_1,
\label{eq:FL_def}
\end{equation}
with the corresponding coefficient function
$C^n_L = C^n_2 - C^n_1$.
A straightforward but tedious calculation gives for each of the moments
\cite{GP76, Blu12}
\begin{subequations}
\label{eq:Mi_def}
\begin{eqnarray}
\label{eq:M1}
M^{(n)}_1(Q^2) &=& \sum_{j = 0}^\infty\,
  \mu^j\,
  \binom{n + j}{j}
  \left(\frac{1}{2}C^{n + 2j}_1
      + \frac{j}{(n + 2j)(n + 2j - 1)} C^{n + 2j}_2
  \right) 
  A_{n + 2j}				\\
\label{eq:M2}
M^{(n)}_2(Q^2) &=& \sum_{j = 0}^\infty\,
  \mu^j\,
  \binom{n+j}{j} \frac{n (n-1)}{(n + 2j)(n + 2j - 1)}\,
  C^{n + 2j}_2\,
  A_{n + 2j}				\\
\label{eq:ML}
M^{(n)}_L(Q^2) &=& \sum_{j = 0}^\infty\,
  \mu^j\,
  \binom{n + j}{j}
  \left(C^{n + 2j}_L + \frac{4j}{(n + 2j)(n + 2j - 1)} C^{n + 2j}_2
  \right)
  A_{n + 2j}				\\
\label{eq:M3}
M^{(n)}_3(Q^2) &=& \sum_{j = 0}^\infty\,
  \mu^j\,
  \binom{n + j}{j} \frac{n}{n + 2j}\,
  C^{n + 2j}_3\,
  A_{n + 2j}				\\
\label{eq:M4}
M^{(n)}_4(Q^2) &=& \sum_{j = 0}^\infty\,
  \mu^j\,
  \binom{n + j}{j} 
  \left( \frac{j(j - 1)}{(n + 2j)(n + 2j - 1)}\, C^{n + 2j}_2
	+ \frac{1}{4} C^{n + 2j}_4
  \right.				\nonumber\\
& & \hspace*{3cm}
  \left. -\ \frac{j}{(n + 2j)(n + 2j - 1)}\, C^{n + 2j}_5
  \right)
  A_{n + 2j}				\\
\label{eq:M5}
M^{(n)}_5(Q^2) &=& \sum_{j = 0}^\infty\,
  \mu^j\,
  \binom{n + j}{j} 
  \frac{n}{n + 2j} 
  \left( - \frac{j}{n + 2j - 1}\, C^{n + 2j}_2
	 + \frac{1}{2} C^{n + 2j}_5
  \right)
  A_{n + 2j}
\end{eqnarray}%
\end{subequations}%
where the binomial symbol $\binom{a}{b} = a!/[b!(a-b)!]$.
Further details of the derivation of Eqs.~(\ref{eq:Mi_def}) are
given in Appendix~A.  Note that the expression for the $M_1^{(n)}$
moment is the same as that in Ref.~\cite{Blu12} once the differences
between the corresponding operator definitions are taken into account
\cite{BluPriv}.

Up to this point the effects of the target mass on the structure
function moments are rigorously derived within the OPE formalism.
To proceed beyond Eqs.~(\ref{eq:Mi_def}) and determine the TMC effects
on the $x$ dependence of the structure functions themselves requires
additional assumptions, which inevitably introduces some model
dependence in the calculation, as we discuss next in the following
section.

% .......................................................................
\subsection{Parton distributions with TMCs}
\label{ssec:PDF}

In the absence of color interactions, the matrix elements $A_{2k}$
in Eq.~(\ref{eq:A2k_def}) should not depend on any scale apart
from the factorization scale.  With this in mind, the products
$C^{2k}_i\, A_{2k}$ in Eq.~(\ref{eq:T}) can be written in terms
of parton distribution functions $f_i$ as
\begin{eqnarray}
C^{2k}_i\, A_{2k} &=& \int_0^1 dy\, y^{2k-1}\, f_i(y),
\label{eq:f_i}
\end{eqnarray}
where for ease of notation we suppress the dependence in $C^{2k}_i$
and $f_i$ on the scale $Q^2$, which arises from perturbative QCD
corrections.
The functions $f_i$ are defined such that in the massless limit
$(\mu \to 0)$ one has
\begin{equation}
\left\{
  F^{(0)}_1,\ F^{(0)}_2,\ F^{(0)}_L,\
  F^{(0)}_3,\ F^{(0)}_4,\ F^{(0)}_5
\right\}
=
\left\{
  \frac{1}{2} f_1,\ x f_2,\ x (f_2 - f_1),\
  f_3,\ \frac{1}{4} f_4,\ \frac{1}{2} f_5
\right\},
\end{equation}
where $F^{(0)}_i \equiv \lim_{\mu \to 0} F_i$ is the massless limit
of the physical structure function $F_i$.
Note that our notation for the parton distribution functions $f_i$
differs from that in Refs.~\cite{GP76, KR04}, whose distributions
effectively correspond to $f_i(x)/x$.

The functions $f_i$ can in principle be identified with the PDFs
measured in deep-inelastic or other high-energy scattering processes.
(For simplicity we omit the flavor dependence of the structure
functions, including their electroweak couplings, which can be
incorporated straightforwardly with the distributions $f_i$.)
Following the derivation of GP \cite{GP76}, the structure functions
at finite $Q^2$ can be inverted using the inverse Mellin transform,
\begin{eqnarray}
\label{eq:Fi_def}
F_i(x,Q^2)
&=& \left\{
    \begin{array}{ll}
	{\displaystyle\frac{1}{2\pi i} \int_{-i\infty}^{i\infty}} dn\,
		x^{-n}\, M^{(n)}_i(Q^2)
	& \hspace{6 pt} \mbox{if}\ \ i=1, 3, 4, 5	\\
	&						\\
        {\displaystyle\frac{1}{2\pi i} \int_{-i\infty}^{i\infty}} dn\,
		x^{-n+1}\, M^{(n)}_i(Q^2)
	& \hspace{5 pt} \mbox{if}\ \ i=2, L.
    \end{array}
    \right.
\end{eqnarray}
Using Eqs.~(\ref{eq:Mi_def}) and (\ref{eq:f_i}), the $x$ dependence of
the structure functions can then be determined in terms of the functions
$f_i$, as outlined in Appendix~\ref{app:sf} \cite{GP76, KR04},
\begin{subequations}
\label{eq:Fi_full}
\begin{eqnarray}
F_1(x,Q^2)
&=& {1 \over 2(1 + \mu \xi^2)}\, f_1(\xi)\
 -\ \mu x^2 {\partial \over \partial x}
    \left( {g_2(\xi) \over 1 + \mu \xi^2} \right),
\label{eq:F1}							\\
F_2(x,Q^2)
&=& x^2 {\partial^2 \over \partial x^2}
    \left( {x g_2(\xi) \over \xi (1 + \mu \xi^2)} \right),
\label{eq:F2}							\\
F_L(x,Q^2)
&=& -{x \over 1 + \mu \xi^2}\, f_1(\xi)\
 +\ 2\mu x^3 {\partial \over \partial x}
    \left( {g_2(\xi) \over 1 + \mu \xi^2} \right)		\nonumber\\
& &
 +\ (1 + 4\mu x^2) x^2 {\partial^2 \over \partial x^2}
    \left( {x g_2(\xi) \over \xi (1 + \mu \xi^2)} \right),
\label{eq:FL} 							\\
F_3(x,Q^2)
&=& -x {\partial \over \partial x}
    \left( {h_3(\xi) \over 1 + \mu \xi^2} \right),
\label{eq:F3}							\\
F_4(x,Q^2)
&=& {1 \over 4(1 + \mu \xi^2)}\, f_4(\xi)\
 +\ \mu x^2 {\partial \over \partial x}
    \left( {g_5(\xi) \over 1 + \mu \xi^2} \right)		\nonumber\\
& &
 +\ \mu^2 x^3 {\partial^2 \over \partial x^2}
    \left( {\xi^2 g_2(\xi) \over 1 - \mu^2 \xi^4} \right),
\label{eq:F4}							\\
F_5(x,Q^2)
&=& -{x \over 2} {\partial \over \partial x}
    \left( {h_5(\xi) \over 1 + \mu \xi^2} \right)
 -\ \mu x^2 {\partial^2 \over \partial x^2}
    \left( {\xi g_2(\xi) \over 1 - \mu^2 \xi^4} \right),
\label{eq:F5}
\end{eqnarray}
\end{subequations}%
where the functions $h_i$ and $g_i$ are given by
\begin{eqnarray}
h_i(\xi) &=& \int_\xi^1 du\, \frac{f_i(u)}{u},
\label{eq:hi}					\\
g_i(\xi) &=& \int_\xi^1 du\, h_i(u).
\label{eq:gi}
\end{eqnarray}
Note that the expression for the $F_4$ structure function in
Ref.~\cite{KR04} contains ``$\xi h_5$'' instead of ``$g_5$''
in the second term of Eq.~(\ref{eq:F4}).
Equations~(\ref{eq:Fi_full}) define the complete set of unpolarized
structure functions in the standard treatment of TMCs in the OPE.
As was noted already in Ref.~\cite{DGP77}, however, the standard
results lead to problems in the limit as $x \to 1$, which we shall
focus on in the remainder of this section.

% .......................................................................
\subsection{Consistency of the standard TMCs?}
\label{ssec:consistency}

When taking the moments of the calculated $x$-dependent structure
functions in the presence of TMCs, one should for consistency
recover the expressions for the moments in Eqs.~(\ref{eq:Mi_def}).
To be specific, we investigate this here for the $F_2$ structure
function, Eq.~(\ref{eq:F2}), but the same arguments can be applied
to all the other structure functions.  From the definition of the
moments in Eq.~(\ref{eq:momentdef}), the $n$-th moment of $F_2$
can be written as
\begin{subequations}
\begin{eqnarray}
M_2^{(n)}(Q^2)
&=& \int_0^1 dx\, x^n \frac{\partial^2}{\partial x^2}
    \left( \frac{x g_2(\xi)}{\xi (1 + \mu \xi^2)} \right)
\label{eq:M2nX}						\\
&=& \left[ x^n \frac{\partial}{\partial x}
           \left( \frac{x g_2(\xi)}{\xi (1 + \mu \xi^2)} \right)
    \right]_{x=0}^1 
 -\ \left[ n x^{n-1}
	   \frac{x g_2(\xi)}{\xi (1 + \mu \xi^2)}
    \right]_{x=0}^1 					\nonumber\\
& &
 +\ n(n-1) \int_0^1 dx\, x^{n-2}
    \frac{x g_2(\xi)}{\xi (1 + \mu \xi^2)},
\end{eqnarray}
\end{subequations}
where integration by parts has been performed twice.  Changing
variables from $x$ to $\xi$, and using the fact that the kinematic
maximum value of $\xi$ is given by $\xi_0$, the moment becomes
\begin{eqnarray}
M_2^{(n)}(Q^2)
&=& \frac{4 \mu^2\xi_0^3}{(1 + \mu \xi_0^2)^3}\, g_2(\xi_0)\
 +\ \frac{1 - \mu \xi_0^2}{(1 + \mu \xi_0^2)^2}
    \left. \frac{\partial g_2(\xi)}{\partial \xi}
    \right|_{\xi=\xi_0}\
 -\ \frac{n}{(1 - \mu^2 \xi_0^4)}\, g_2(\xi_0)		\nonumber\\
& &
 +\ n(n-1) \sum_{j=0}^\infty\, \mu^j\, \binom{n+j}{j}\,
    \int_0^{\xi_0} d\xi\, \xi^{n+2j-2}\, g_2(\xi),
\label{eq:M2GP}
\end{eqnarray}
where we have also used $dx/d\xi = (1+\mu \xi^2)/(1-\mu \xi^2)^2$,
together with the relation
\begin{equation}
\frac{1}{(1-\mu \xi^2)^{n+1}}
= \sum_{j=0}^\infty\, \mu^j\, \binom{n + j}{j}\, \xi^{2j}.
\label{eq:binomial}
\end{equation}

Now, consider the last term in Eq.~(\ref{eq:M2GP}) involving the
integral of the function $g_2(\xi)$.  From the definition of the
parton distributions in Eq.~(\ref{eq:f_i}), one can write
\begin{eqnarray}
\frac{1}{(n+2j)(n+2j-1)}\, C_2^{n+2j}\, A_{n+2j}
&=& \int_0^{\xi_0} d\xi\, \xi^{n+2j-2}\, g_2(\xi)\
 +\ \int_{\xi_0}^1 d\xi\, \xi^{n+2j-2}\, g_2(\xi).
\label{eq:AnNew}
\end{eqnarray}
However, because the function $f_2$ (and hence its integrals as in
Eqs.~(\ref{eq:hi}) and (\ref{eq:gi})) has no reason to vanish in the
region $\xi_0 < \xi < 1$, the second term in Eq.~(\ref{eq:AnNew})
is in general nonzero.  The same is true for the first three terms
in Eq.~(\ref{eq:M2GP}), and as a consequence one does not recover
exactly the original expression, Eq.~(\ref{eq:M2}).

On the other hand, if the parton distributions were to vanish in the
region $\xi_0 < \xi < 1$, the moments would have to depend on $\xi_0$,
\begin{equation}
C_i^n\, A_n(\xi_0)
= \int_0^1 d\xi\, \xi^{n-1}\, f_i(\xi;\xi_0)\ \
\Longrightarrow\ \
  \frac{dA_n(\xi_0)}{d\xi_0}
= \int_0^1 d\xi\, \xi^n\, \frac{df_i(\xi;\xi_0)}{d\xi_0}\
\neq\ 0,
\end{equation}
where we explicitly label the dependence of the functions $f_i$
on $\xi$ and $\xi_0$.
This result suggests two immediate problems:
(i) universal (process-independent) parton distributions would
no longer exist at finite $Q^2$; and
(ii) the separation between short and long distances on the
light-cone, as embodied in the OPE, would no longer be possible.

If the condition that the structure functions vanish for $\xi > \xi_0$
is not imposed, one is then faced with the prospect of energy-momentum
not being conserved.  In fact, if the upper limit of integration in
Eq.~(\ref{eq:M2nX}) were extended from $x = 1$ to $x = 1/(1-\mu)$,
the first three terms of Eq.~(\ref{eq:M2GP}) would be identically
zero, and extending the integration in the fourth term to $\xi=1$,
Eq.~(\ref{eq:M2}) would be recovered.  Consequently, the consistency of
the GP prescription \cite{GP76}, and in most subsequent TMC treatments,
requires the violation of energy-momentum conservation.  It thus appears
a general consequence of defining parton distributions at finite $Q^2$
in the presence of TMCs that one must choose between two less than
ideal options: either keeping a universal parton distribution and
violating energy-momentum conservation, or conserving energy and
momentum but working with process-dependent distributions.

\begin{figure}
\includegraphics[width=11cm]{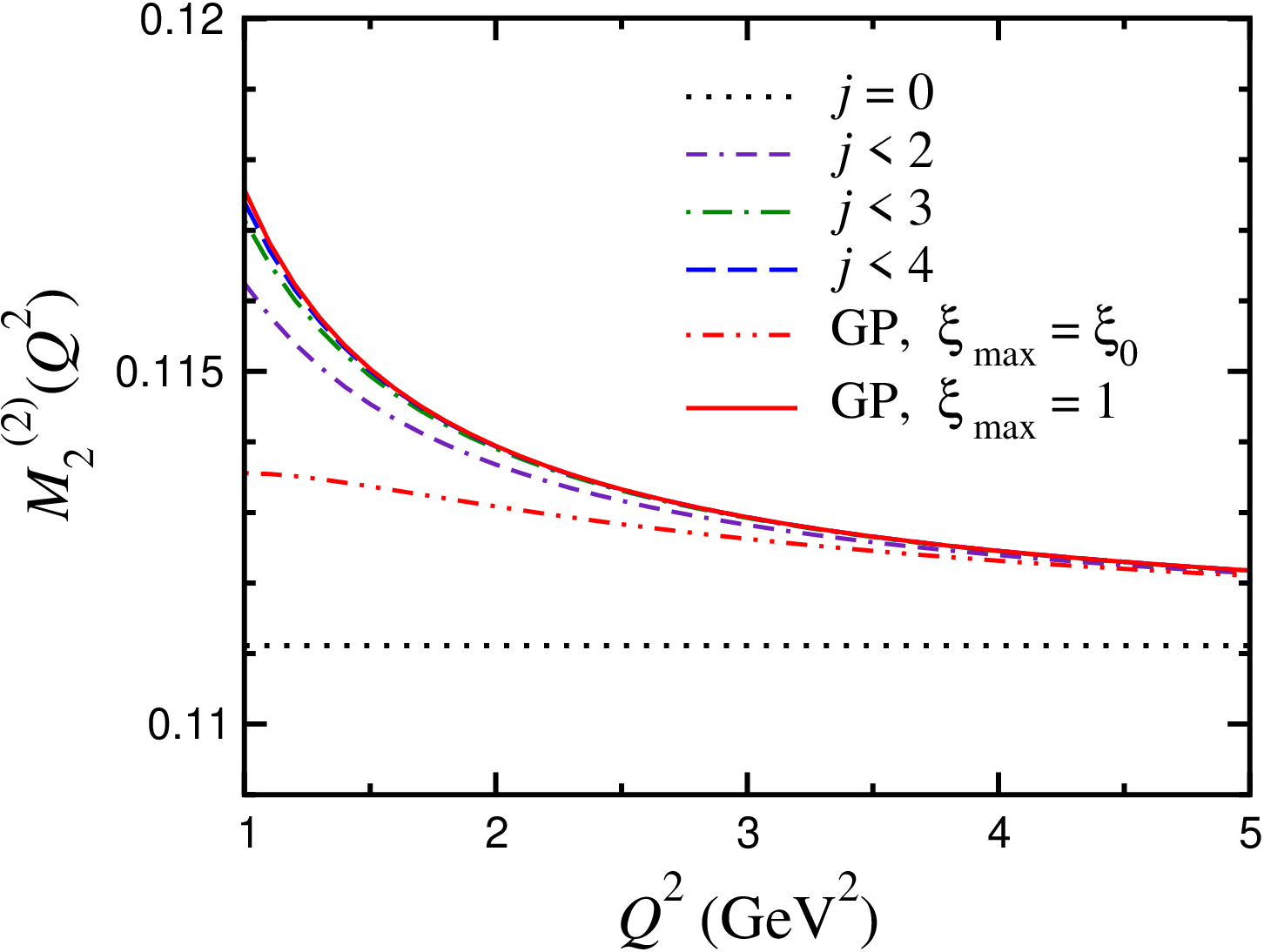}
\caption{$n=2$ moments of the $F_2$ structure function, illustrating
	the convergence of the series in Eq.~(\ref{eq:M2}) for $j=0$
	(dotted), $j<2$ (dot-dash-dashed), $j<3$ (dot-dashed) and
	$j<4$ (dashed), compared with the standard TMC result from
	GP \cite{GP76} using Eq.~(\ref{eq:F2}) with the upper limit
	of integration $\xi_{\rm max}=\xi_0$ (\ref{eq:M2GP})
	(dot-dot-dashed) and $\xi_{\rm max}=1$ (solid).}
\label{fig:mom}
\end{figure}

We can assess the numerical significance of the $\xi > \xi_0$ region
by evaluating the lowest ($n=2$) moment $M_2^{(2)}$ using a simple
form for the parton distribution,
\begin{eqnarray}
x f_2(x) &=& {35 \over 32}\, \sqrt{x}\, (1-x)^3,
\label{eq:f2_ex}
\end{eqnarray}
chosen to approximately reproduce a typical valence quark distribution,
normalized such that $\int_0^1 dx\, f_2(x) = 1$.
The moment shown in Fig.~\ref{fig:mom} is computed from
Eq.~(\ref{eq:M2}) for several values of $j$ (from the leading term
only, $j=0$, up to the inclusion of the first four terms, $j<4$),
and is compared with directly integrating $F_2(x,Q^2)$ over $x$
from 0 to 1 (or equivalently up to $\xi = \xi_{\rm max} = \xi_0$),
using Eqs.~(\ref{eq:F2}).
As noted above, this procedure does not recover the formal result
for the moment, Eq.~(\ref{eq:M2}), which is reflected in the
nonconvergence of the moments with increasing $j$ to the standard
TMC result from GP \cite{GP76} in Eq.~(\ref{eq:F2})
(dot-dot-dashed curve in Fig.~\ref{fig:mom}).
On the other hand, if the missing integration range as expressed
in the second term of Eq.~(\ref{eq:AnNew}) is kept, the convergence
of the moments is recovered (solid curve in Fig.~\ref{fig:mom}),
although at the expense of effectively integrating beyond $x=1$
(to $\xi_{\rm max}=1$).

The problem encountered here is at the core of the parton
interpretation of the matrix elements $A_n$.  The approach of GP
attempts to maintain a partonic interpretation at finite $Q^2$ by
introducing a new scaling variable $\xi$ \cite{Nac73, Gre71}.
However, as shown in Eq.~(\ref{eq:M2GP}), this leads to
inconsistencies in the extracted $x$ dependence of the structure
functions and their moments.

A possible way to avoid the problematic $\xi \sim \xi_0$ region is,
ironically, to not introduce the Nachtmann scaling variable $\xi$ in
the first place.  This can be realized by performing the inversion
of the moments order by order in $\mu$, rather than summing over all
powers of $\mu$ during the inversion.  As we shall see in the next
section, this allows us to work with universal twist-2 distribution
functions, while simultaneously preserving energy-momentum conservation.
The only drawback of this approach is that the region of $x$ and $Q^2$
where parton distributions can be formulated consistently in the
presence of TMCs will be somewhat restricted.

%%%%%%%%%%%%%%%%%%%%%%%%%%%%%%%%%%%%%%%%%%%%%%%%%%%%%%%%%%%%%%%%%%%%%%%%%
\section{Series expansion of inverted moments}
\label{sec:series}

In the course of inverting the moments to obtain the structure
functions, the binomial theorem is used to perform the integration
by absorbing combinatorial factors involving the integration variable
$n$ (see Appendix~\ref{app:sf}).  Instead of this standard procedure,
in this section we describe how the moments can be inverted term by term
by absorbing the combinatorial factor into derivatives, which gives
rise to novel series expansions for each of the structure functions.
We illustrate this procedure for the case of the $F_2$ structure
function, with the derivation of the other structure functions
following similarly.

For the $j$-th term in the series expansion for the $F_2$ moment in
Eq.~(\ref{eq:M1}), $M_{2,j}^{(n)}$, the contribution to the structure
function is given by the inverse Mellin transform
\begin{eqnarray}
F_{2,j}(x,Q^2)
&=& \frac{1}{2\pi i}
    \int_{-i\infty}^{i\infty} dn\, x^{1-n}\, M_{2,j}^{(n)}(Q^2).
\end{eqnarray}
Using integration by parts to write
\begin{eqnarray}
C^{n+2j}_2\, A_{n+2j}
&=& (n+2j)(n+2j-1) \int_0^1 dy\, y^{n+2j-2}\, g_2(y,Q^2),
\label{eq:parts}
\end{eqnarray}
the contribution to $F_{2,j}$ can then be expressed in the form
\begin{eqnarray}
F_{2,j}(x,Q^2)
&=& \mu^j
    \frac{1}{2\pi i}
    \int_{-i\infty}^{i\infty} dn\,
    \int_0^1 dy\,
    \frac{(n + j)!}{j!(n - 2)!}\,
    x^{1 - n}\, y^{n + 2j - 2}\, g_2(y).
\label{eq:generic}
\end{eqnarray}
Next, we can observe that 
\begin{equation}
  {(n + j)! \over (n - 2)!}\, x^{-n + 1}\,
= (-x)^{2 + j}\,
  {\partial^{2 + j} \over \partial x^{2 + j}}\, x^{-n + 1}
\label{eq:deriv_x}
\end{equation}
for all $n$ on the imaginary axis except the origin, so that
\begin{eqnarray}
F_{2,j}(x,Q^2)
%	&= \left( \frac{M^2}{Q^2} \right)^j
%	   \frac{(-x)^{2 + j}}{j!} 
%	   \frac{\partial^{2 + j}}{\partial x^{2 + j}}
%	   \frac{1}{2\pi i}
%	   \int_{-i\infty}^{i\infty} dn \int_0^1 dy\, x^{-n + 1} 
%	   y^{n + 2j - 2} g_2(y)				\\
&=& \mu^j
    \frac{(-x)^{2 + j}}{j!} 
    \frac{\partial^{2 + j}}{\partial x^{2 + j}}
    \int_0^1 dy\, x\, y^{2j - 2}\, g_2(y)\,
    \frac{1}{2\pi i}
    \int_{-i\infty}^{i\infty} dn\,
    \left( \frac{y}{x} \right)^n.
%
% &= \left( \frac{M^2}{Q^2} \right)^j
%	   \frac{(-x)^{2 + j}}{j!} 
%	   \frac{\partial^{2 + j}}{\partial x^{2 + j}}
%	   \int_0^1 dy\, x y^{2j - 2} g_2(y)
%	   \delta\left(\ln\frac{y}{x}\right)			\\
%	&= \left(\frac{M^2}{Q^2}\right)^j \frac{(-x)^{2 + j}}{j!} 
%	   \frac{\partial^{2 + j}}{\partial x^{2 + j}} \int_0^1 x 
%	   y^{2j - 2} g_2(y) x \delta(y - x) \, dy \\
%	&= \left(\frac{M^2}{Q^2}\right)^j \frac{(-x)^{2 + j}}{j!} 
%	   \frac{\partial^{2 + j}}{\partial x^{2 + j}} 
%	   \left(x^{2j} g_2(x) \right).
\label{eq:F2j_def}
\end{eqnarray}
Finally, making use of the $\delta$-function representation in
Eq.~(\ref{eq:delta-fn})
%
% \begin{eqnarray}
% \delta(u)
% &=& \frac{1}{2\pi i} \int_{-i\infty}^{i\infty} dn\, u^{n-1},
% \label{eq:delta-fn}
% \end{eqnarray}
%
we arrive at the desired result,
\begin{eqnarray}
F_{2,j}(x,Q^2)
&=& \frac{(-x)^{2 + j}}{j!}\, \mu^j\,
    \frac{\partial^{2 + j}}{\partial x^{2 + j}}
    \left[ x^{2j}\, g_2(x) \right].
\label{eq:F2j}
\end{eqnarray}

This result can also be obtained by noting that, instead of
Eq.~(\ref{eq:deriv_x}), we can write
\begin{equation}
  \frac{(n + j)!}{(n - 2)!}\, y^{n + 2j - 2}
= y^{2j}\, \frac{\partial^{2 + j}}{\partial y^{2 + j}}\, y^{n + j}.
\end{equation}
Substituting this into Eq.~(\ref{eq:generic}) then leads to
\begin{eqnarray}
F_{2,j}(x,Q^2)
&=& \mu^j\, \frac{1}{j!}\, \frac{1}{2\pi i}\,
    \int_{-i\infty}^{i\infty} dn\,
    \int_0^1 dy\, x^{-n + 1} y^{2j} 
    \frac{\partial^{2 + j}}{\partial y^{2 + j}}
    \left[ y^{n + j} g_2(y) \right]		\nonumber\\
%	&= \left(\frac{M^2}{Q^2}\right)^j \frac{1}{j!} \int_0^1 x 
%	   y^{2j} g_2(y) \frac{\partial^{2 + j}}{\partial y^{2 + j}} 
%	   \left( y^j \frac{1}{2\pi i} \int_{-i\infty}^{i\infty} 
%	   \left(\frac{y}{x}\right)^n \right)  \\
&=& \mu^j\, \frac{x^2}{j!}\,
    \int_0^1 dy\,
    y^{2j}\, g_2(y)
    \frac{\partial^{2 + j}}{\partial y^{2 + j}} 
    \left[ y^j \delta(y - x) \right],
\label{eq:dist}
\end{eqnarray}
using again the $\delta$-function representation (\ref{eq:delta-fn}).
%
% Now, integrating by parts the action of a distribution $F$ on a
% function $\phi$,
% \begin{equation}
% \int dz\, F'(z)\, \phi(z) = -\!\int dz\, F(z) \phi'(z),
% \end{equation}
% where the distribution $F'$ is defined as the derivative of $F$,
%
Now, because the $\delta$-function is a distribution, one defines its
derivative (in analogy with integration by parts of regular functions) as
\begin{equation}
\int dz\, \delta(z)\, \phi'(z) = -\int dz\, \delta'(z)\, \phi(z)
\end{equation}
for a given function $\phi$.  Applying this definition $(2 + j)$
times to Eq.~(\ref{eq:dist}), we find
\begin{align}
F_{2,j}(x,Q^2)
&= \mu^j\,
   \frac{(-1)^{2 + j}}{j!}\, x^2
   \int_0^1 dy\,
   \frac{\partial^{2 + j}}{\partial y^{2 + j}}
   \left[ y^{2j} g_2(y) \right]\, y^j\, \delta(y - x),
%
%	&= \left(\frac{M^2}{Q^2}\right)^j \frac{(-x)^{2 + j}}{j!}
%	   \frac{\partial^{2 + j}}{\partial x^{2 + j}}
%	   \left(x^{2j} g_2(x) \right).
\end{align}
which gives a result identical to that in Eq.~(\ref{eq:F2j}).

Either of these two methods may be applied to the moments of the
other structure functions to obtain the complete expressions for
the unpolarized TMC structure functions in terms of power series
in $M^2/Q^2$, summing over all values of $j$,
\begin{subequations}
\label{eq:Fi_ser}
\begin{eqnarray}
F_1(x,Q^2) &=& x \sum_{j=0}^\infty\, \mu^j\,
	\frac{(-x)^j}{j!}
	\frac{\partial^j}{\partial x^j} 
	  \left[ x^{2j - 2}
		 \left( {1 \over 2} x f_1(x) + j g_2(x) \right)
	  \right],
\label{eq:F1ser}				\\
F_2(x,Q^2) &=& x^2 \sum_{j=0}^\infty\, \mu^j\,
	\frac{(-x)^j}{j!}
	\frac{\partial^{2 + j}}{\partial x^{2 + j}}
	  \left[ x^{2j} g_2(x)
	  \right],
\label{eq:F2ser}				\\
F_L(x,Q^2) &=& x^2 \sum_{j=0}^\infty\, \mu^j\,
	\frac{(-x)^j}{j!}
	\frac{\partial^j}{\partial x^j} 
	  \left[ x^{2j - 2}
		 \left( xf_2(x) - xf_1(x) + 4 j g_2(x) \right)
	  \right],
\label{eq:FLser}				\\
F_3(x,Q^2) &=& \sum_{j=0}^\infty\, \mu^j\,
	\frac{(-x)^{1 + j}}{j!}
	\frac{\partial^{1 + j}}{\partial x^{1 + j}}
	  \left[ x^{2j} h_3(x)
	  \right],
\label{eq:F3ser}				\\
F_4(x,Q^2) &=& x \sum_{j=0}^\infty\, \mu^j\,
	\frac{(-x)^j}{j!}
	\frac{\partial^j}{\partial x^j}
	  \left[ x^{2j - 2}
		 \left( j(j - 1) g_2(x) + {1 \over 4}x f_4(x) - j g_5(x)
		 \right)
	  \right],
\label{eq:F4ser}				\\
F_5(x,Q^2) &=& \sum_{j=0}^\infty\, \mu^j\,
	\frac{(-x)^{1 + j}}{j!}
	\frac{\partial^{1 + j}}{\partial x^{1 + j}}
	  \left[ x^{2j - 1}
		 \left( -j g_2(x) + \frac{1}{2} x h_5(x) \right)
	  \right].
\label{eq:F5ser}
\end{eqnarray}
\end{subequations}%
Note that the Nachtmann variable $\xi$ does not enter in
Eqs.~(\ref{eq:Fi_ser}), and all the functions $f_i$, $g_i$ and $h_i$
are expressed as functions of $x$ only.  As required, the $j=0$ term
in the expansion of $F_i$ is simply the massless limit structure
function, $F^{(0)}_i$.
%
% For any nonzero $j$ one can verify these results by directly expanding
% the structure functions in Eqs.~(\ref{eq:Fi_full}), with $\xi$ expanded
% about $x$ and $f_i(\xi)$ about $f(x)$.

The advantage of this formulation is that it explicitly avoids the
problems encountered with the consistency of the inversion in the GP
approach discussed in Sec.~\ref{ssec:consistency}.  Indeed, direct
integration of the structure functions in (\ref{eq:Fi_ser}) leads to
the correct expressions for the moments in Eqs.~(\ref{eq:Mi_def}).

\begin{figure}
\includegraphics[width=11cm]{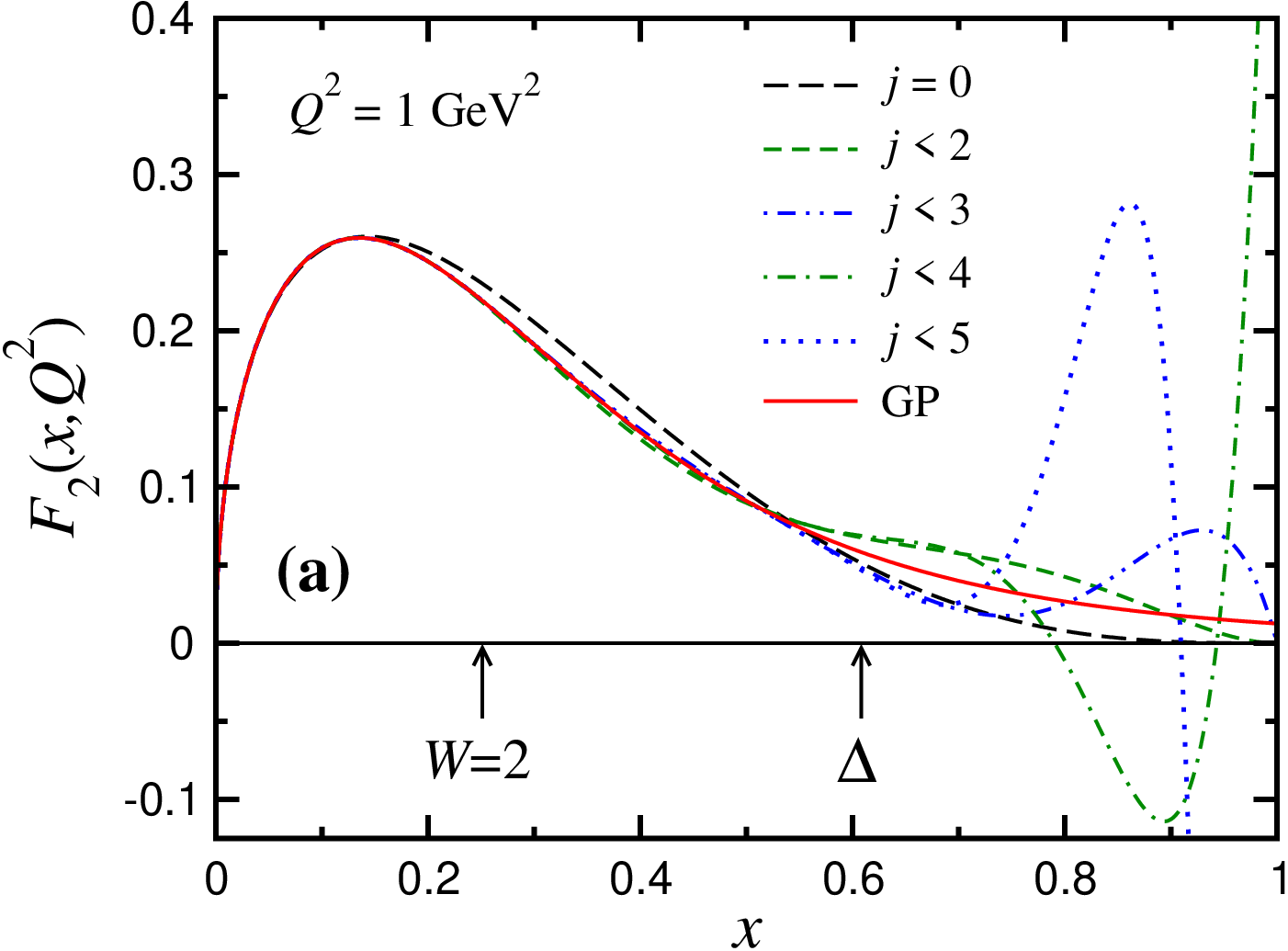}\vspace*{1.5cm}
\includegraphics[width=11cm]{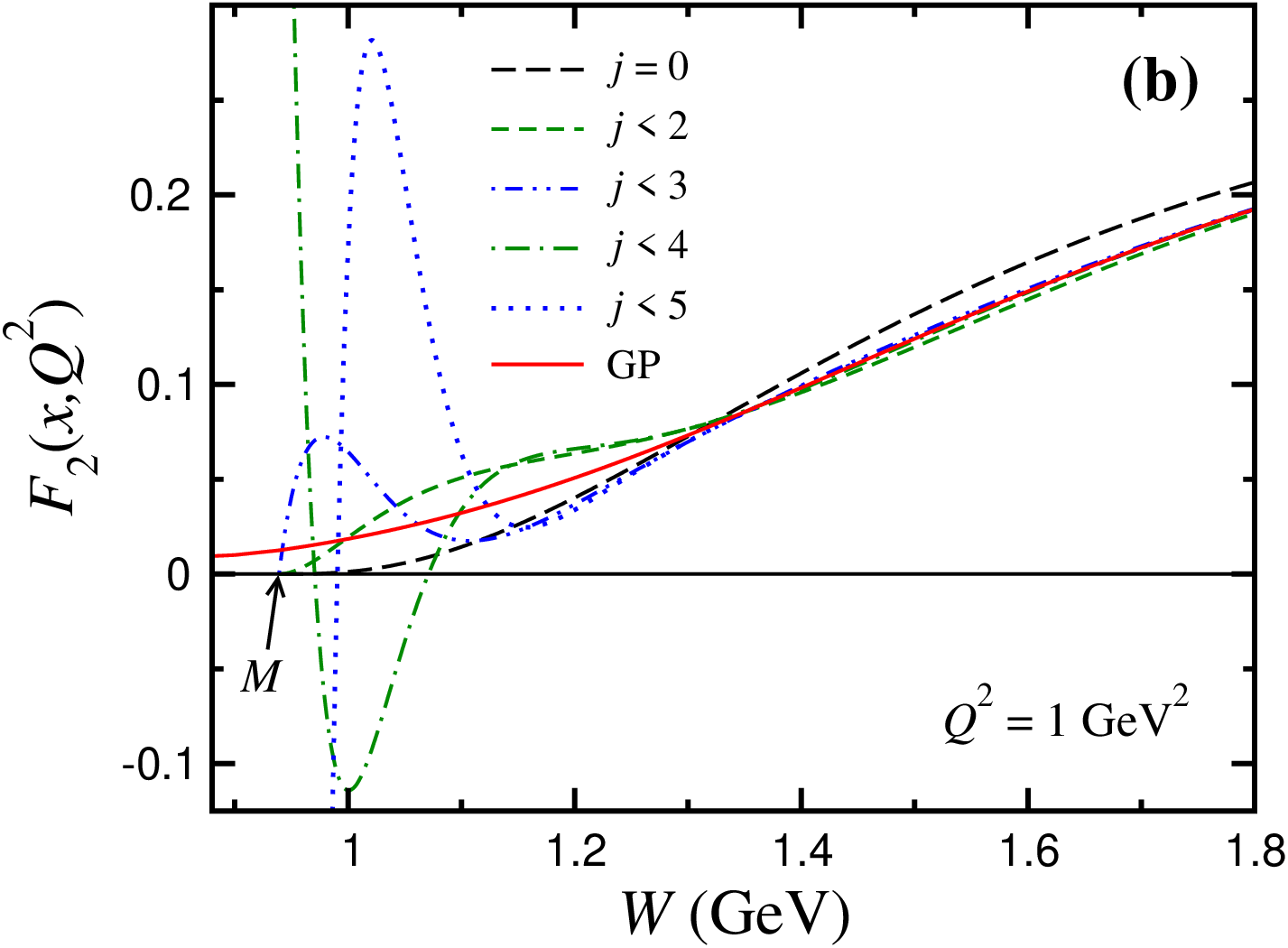}
\caption{Target mass corrected $F_2$ structure function at
	$Q^2=1$~GeV$^2$ from Eq.~(\ref{eq:F2ser}), showing the
	convergence with increasing $j$, and compared with the
	standard TMC result from GP \cite{GP76} using Eq.~(\ref{eq:F2}),
	shown as a function of (a) Bjorken-$x$, and (b) the hadronic
	final state mass $W$.  The arrows in (a) indicate the locations
	of the resonance region ($W=2$~GeV) and the $\Delta$ resonance,
	while the arrow in (b) denotes the elastic limit, $W=M$.}
\label{fig:F2Q1}
\end{figure}

To examine the convergence of the series in Eqs.~(\ref{eq:Fi_ser}),
we show in Fig.~\ref{fig:F2Q1}(a) the first few terms in the expansion
of $F_2$, starting with the leading order, $j=0$, term and up to
the first five terms in the series, $j<5$.
For illustration, we use the simple massless limit function in
Eq.~(\ref{eq:f2_ex}), and compare the result with the standard
TMC calculation from GP \cite{GP76} in Eq.~(\ref{eq:F2}).
The results show that the convergence at $Q^2 = 1$~GeV$^2$ is
fairly rapid for $x \lesssim 0.5$, with just the first two or
three terms already giving a target mass corrected function that
does not change noticeably with inclusion of higher orders.

It is noteworthy that in this range one is already well within the
nucleon resonance region, traditionally taken to be $W < 2$~GeV,
from which data are typically excluded in global PDF analyses.
This can be more clearly seen in Fig.~\ref{fig:F2Q1}(b),
where the structure function is shown as a function of $W$.
The convergence of the TMCs is well under control down to values
as low as $W \approx 1.3$~GeV, just above the peak of the first
resonance region dominated by the $\Delta(1232)$ resonance.
At smaller $W$, or higher $x$, the higher order terms display
oscillatory behavior as one approaches the nucleon elastic point,
$x=1$ (or $W=M$).  For the particular form of $f_2$ chosen in
Eq.~(\ref{eq:f2_ex}), $xf_2 \sim (1-x)^3$, the first three terms in
the series ($j < 3$) vanish as $x \to 1$, while the contributions
for $j \geq 3$ diverge at $x = 1$.  The target mass corrected function
from Eq.~(\ref{eq:F2}) (labeled ``GP'' in Fig.~\ref{fig:F2Q1}) is
finite at $x=1$ and indeed extends into the unphysical region $W < M$.

\begin{figure}
\includegraphics[width=11cm]{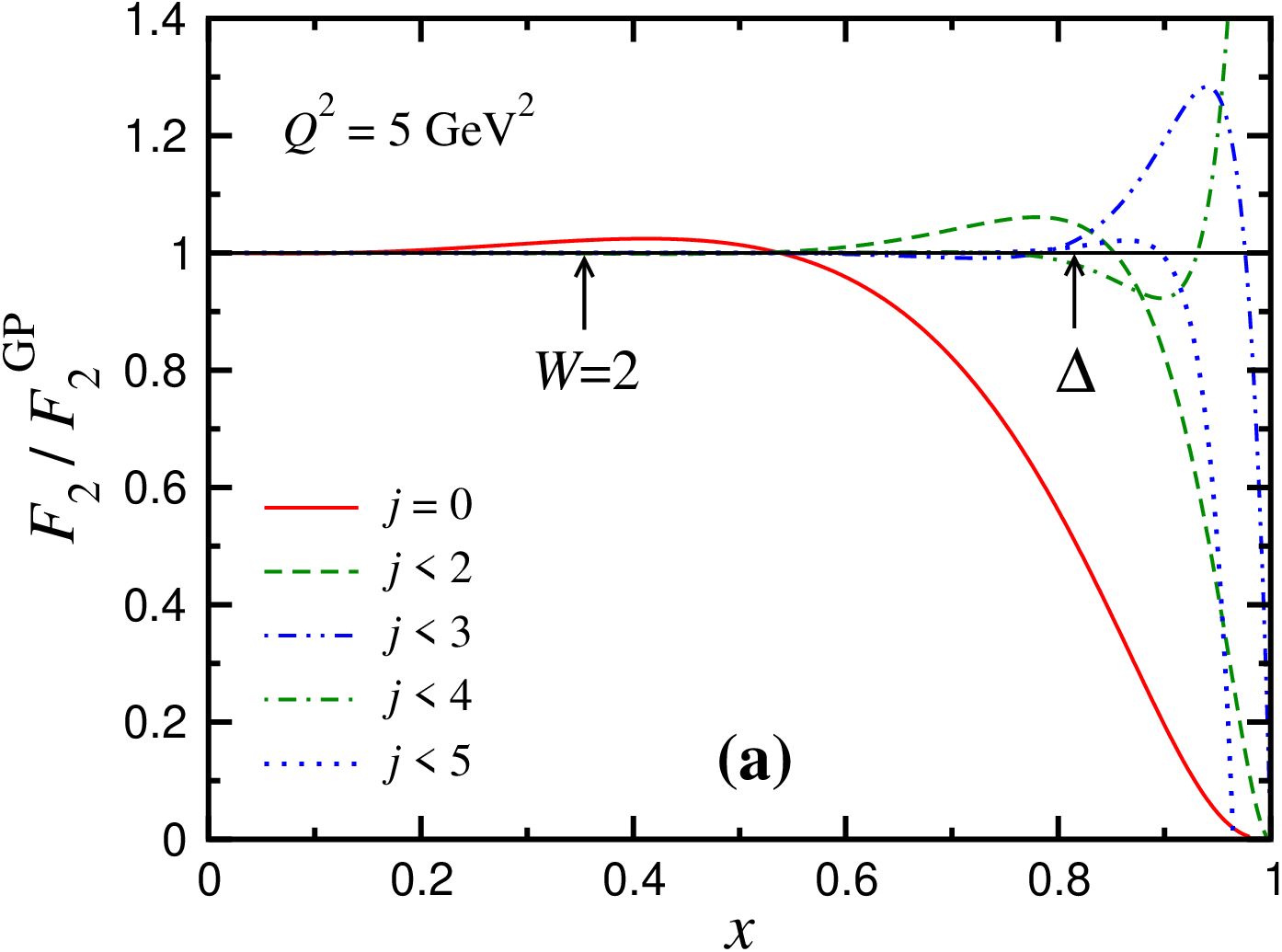}\vspace*{1.5cm}
\includegraphics[width=11cm]{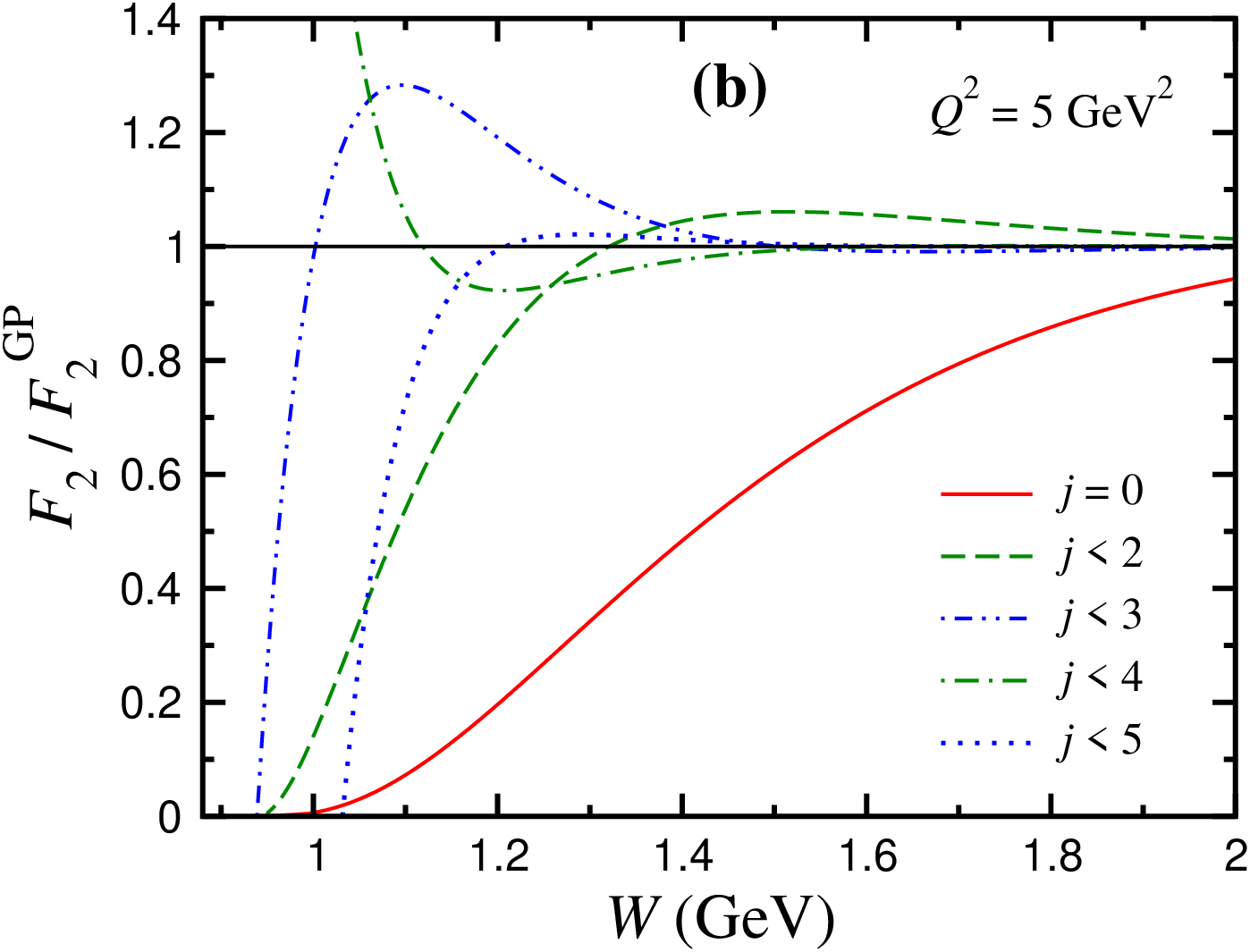}
\caption{Ratio of target mass corrected $F_2$ structure function
	at $Q^2=5$~GeV$^2$ from Eq.~(\ref{eq:F2ser}) for various
	$j$ ($j=0$ up to $j<5$) to the standard TMC result from GP
	\cite{GP76} using Eq.~(\ref{eq:F2}), shown as a function of
	(a) Bjorken-$x$, and (b) the hadronic final state mass $W$.
	The arrows in (a) indicate the locations of the resonance
	region ($W=2$~GeV) and the $\Delta$ resonance.}
\label{fig:F2Q5}
\end{figure}

The large-$x$ oscillatory behavior is significantly dampened by the
time one reaches $Q^2 = 5$~GeV$^2$, with the first three terms ($j<3$)
converging well up to $x \approx 0.8$, as shown in Fig.~\ref{fig:F2Q5}(a)
for the ratio of $F_2$ to the GP target mass corrected function,
Eq.~(\ref{eq:F2}).
At this $Q^2$ this corresponds to values of $W \gtrsim 1.4$~GeV,
illustrated in Fig.~\ref{fig:F2Q5}(b), which again is well outside of
the range where DIS data are typically used in global PDF analyses.
Note that the vanishing of ratio of the leading order term, $j=0$,
to the full GP result as $x \to 1$ reflects the nonzero value of
the GP TMC function at $x \geq 1$.

\begin{figure}
\includegraphics[width=11cm]{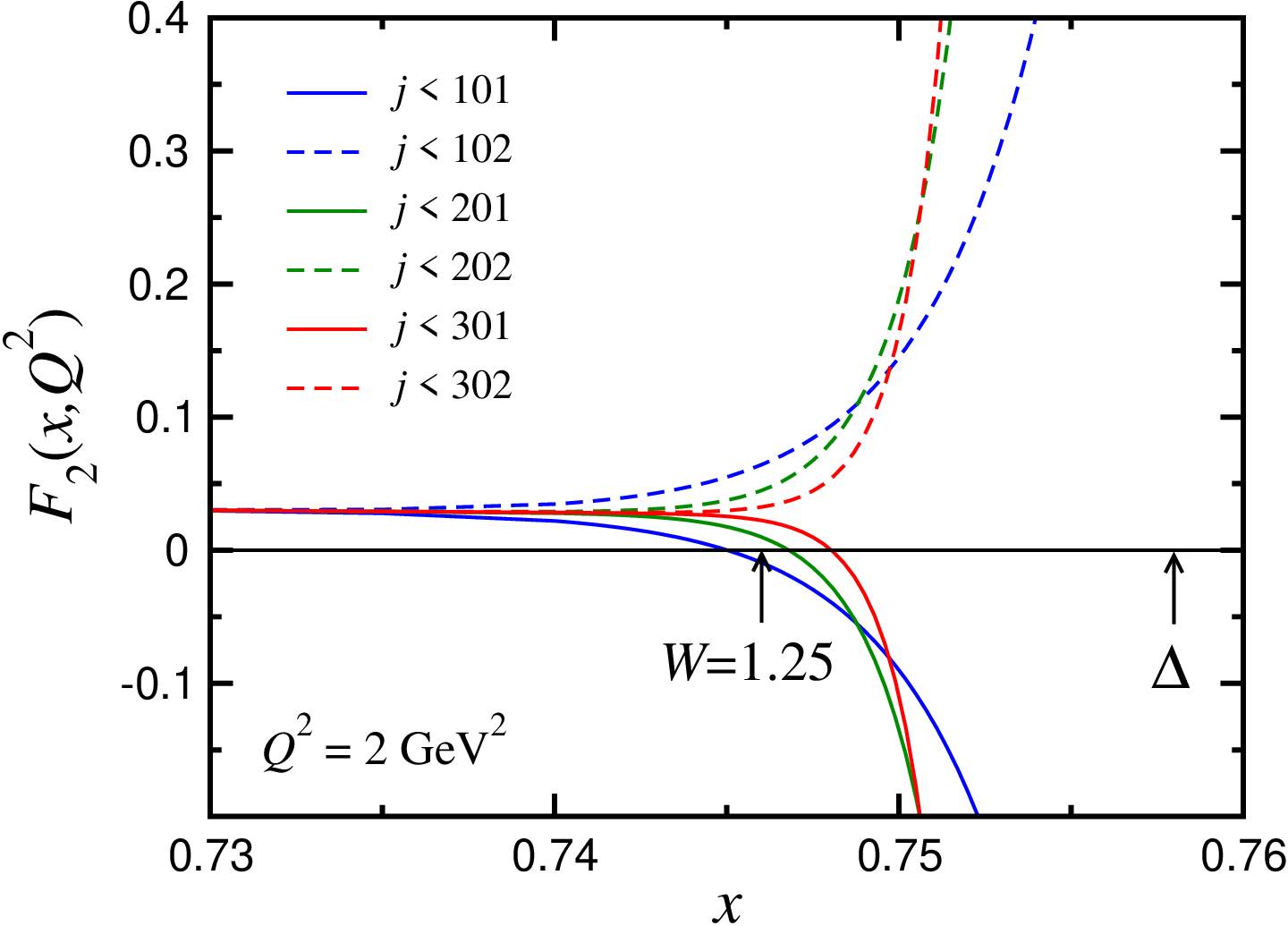}
\caption{Convergence of the series expansion for the target mass
	corrected $F_2$ structure function for large values of $j$
	($j<101, 102, 201, 202, 301$ and 302), at $Q^2 = 2$~GeV$^2$.
	The arrows indicate the values of $x$ at which the final
	state mass corresponds to the $\Delta$ resonance and to
	$W = 1.25$~GeV.  Note the limited $x$ range on the ordinate.}
\label{fig:F2hij}
\end{figure}

One may ask whether the series converges over the entire range of $x$
at finite $Q^2$ if a sufficiently large number of terms is included
in the sum over $j$.  For the trial distribution (\ref{eq:f2_ex})
used here, Fig.~\ref{fig:F2hij} shows the result of summing up to
$\approx 100$, 200 or 300 terms for $Q^2 = 2$~GeV$^2$.
For $x \lesssim 0.73$ (or $W \gtrsim 1.27$~GeV) fairly good
convergence is observed, while for $x \gtrsim 0.74$
(or $W \lesssim 1.25$~GeV) the addition of a large number of terms
is needed to push the convergence of the target mass corrected
function to significantly higher $x$ values.
It is interesting to observe that the inclusion of additional
odd or even terms in $j$ gives alternating negative and positive
divergent behaviors, respectively.
The systematics of this convergence with $Q^2$ and $W$ will be
discussed in more detail in Ref.~\cite{Bro12}.  What is clear,
however, is that a severe limitation on the computation of TMC
exists at very high $x$ for small values of $Q^2$, although this
occurs deep in the resonance region at low $W$.

%%%%%%%%%%%%%%%%%%%%%%%%%%%%%%%%%%%%%%%%%%%%%%%%%%%%%%%%%%%%%%%%%%%%%%%%%
\section{Conclusion}
\label{sec:conc}

The problem of target mass corrections to deep-inelastic structure
functions is almost as old as the theory of QCD itself.  With the
focus of most structure function analyses being on the perturbative
region where subleading $1/Q^2$ effects can be neglected, the
study of TMCs remained largely dormant for several decades.
The advent of new, high-precision data in the resonance--scaling
transition region at high $x$ and low $Q^2$ has brought the problem
of TMCs back to the fore, giving rise to greater urgency to the
need for resolution of the remaining open issues with respect to
their implementation.

In this paper we have sought to illustrate the inherent problem with
the standard TMC formulation, already evident in the pioneering work of
Georgi and Politzer \cite{GP76}, in the treatment of the $x \approx 1$
region and the inversion of the structure functions from their moments.
In particular, we have critically analyzed the definition of PDFs in
the presence of TMCs, and discussed the consequences of the violation
of energy and momentum conservation in the standard TMC analysis.
Historically it has been argued \cite{DGP77, DGPannals} that the
problem in the threshold region exists because at low $Q^2$ the
higher twist contributions cannot be neglected.  We do not disagree
that higher twists are essential for describing low energy structure
function data; we believe, however, that one ought to maintain
consistency of leading twist functions at any $x$, regardless of
how large the higher twists may be at a given $Q^2$.

We contend that the introduction of the Nachtmann variable $\xi$,
which appears naturally in the standard TMC implementation, does not
lead to self-consistent parton distributions that are valid at all $x$.
In fact, our analysis suggests that strictly speaking PDFs cannot be
defined consistently at any finite $Q^2$ when the mass of the target
is incorporated.  What is feasible, however, is to compute the $x$
dependence of the TMC structure functions in terms of standard PDFs
as a series whose convergence can be studied as a function of $x$
and $Q^2$.

To this end, we have derived formulas for the entire set of unpolarized
structure functions, as a series in $M^2/Q^2$, involving PDFs and their
derivatives.  The virtue of this approach is that the resulting TMC
functions can be consistently inverted from their moments, without ever
encountering unphysical regions of kinematics or violating energy and
momentum conservation.  Moreover, it allows us to systematically study
the regions of $x$ and $Q^2$ where TMCs can be reliably applied.
Using a simple trial function, we have illustrated the convergence of
our scheme numerically for the case of the $F_2$ structure function.
Rapid convergence is observed for most of the range of $x$, with the
first two or three terms saturating the sum well into the nucleon
resonance region.  For $Q^2$ values as low as 1~GeV$^2$, we find that
the convergence of the TMC series expansion is under control down to
$W \approx 1.3$~GeV, which is almost in the vicinity of the $\Delta$
resonance peak.

At smaller $W$, or higher $x$ for fixed $Q^2$, rapid oscillations
ensue as one approaches the elastic scattering limit, and beyond
$W \approx 1.3$~GeV it becomes prohibitively difficult to tame these
with a finite number of higher order terms.  At $Q^2 = 2$~GeV$^2$,
for example, even summing over $\sim 300$ terms allows for a smooth
TMC function up to $x \approx 0.73$ (or $W \approx 1.27$~GeV).
Fortunately, such low $W$ values are well outside the range typically
encountered in perturbative QCD analyses of DIS data, and in practice
will not pose any serious restrictions.

Our results therefore lend greater support to global PDF fits which
incorporate low-$W$ data, currently down to $W^2 = 3$~GeV$^2$ in some
analyses \cite{ABKM, CJ10, CJ11}, but with more ambitions plans to
extend the range further into the traditional resonance region.
Provided higher twist and other subleading corrections are tractable,
our method for accounting for TMCs will introduce only minimum
theoretical uncertainty into global analyses.  A more detailed
discussion of the utility of the present approach, as well as its
application to spin-dependent structure functions, will be presented
in a forthcoming publication \cite{Bro12}.

%%%%%%%%%%%%%%%%%%%%%%%%%%%%%%%%%%%%%%%%%%%%%%%%%%%%%%%%%%%%%%%%%%%%%%%%%
\acknowledgments

We thank A.~Accardi, J.~Bl\"umlein and P.~Jimenez-Delgado for helpful
discussions.
This work was supported by the DOE contract No. DE-AC05-06OR23177,
under which Jefferson Science Associates, LLC operates Jefferson Lab.
F.M.S. is supported by CNPq 307675/2009-2.
M.B. acknowledges support from NSF and DoD's ASSURE program which
funded an REU internship at ODU/Jefferson Lab.

\newpage
\appendix
%%%%%%%%%%%%%%%%%%%%%%%%%%%%%%%%%%%%%%%%%%%%%%%%%%%%%%%%%%%%%%%%%%%%%%%%%
\section{Derivation of structure function moments}
\label{app:mom}

In this appendix we illustrate the derivation of moments of structure
functions in the presence of TMCs, using as an example the $F_1$
structure function (the $F_1$ case contains some more general
features that are not present in the $F_2$ derivation discussed
in Secs.~\ref{sec:ope} and \ref{sec:series}).  The results for
the other structure functions follow in a similar manner.

We begin by finding $T_1$, the coefficient of $-g^{\mu \nu}$ in
Eq.~(\ref{eq:T}), which has contributions from both the $C^{2k}_1$
and $C^{2k}_2$ terms.  For the $C^{2k}_1$ term, for fixed
$k \in \mathbb{N}, j \in \{0, \cdots, k \}$, and each term of
$\{g \cdots g \hspace{1 pt} p \cdots p \}_{k, j}$, a total of
$j$\, $q_{\mu_i}$ factors will have their indices raised by $j$\,
$g^{\mu_i \mu_l}$ metric tensors.  The result will then contract
with $j$ $q_{\mu_l}$'s to give a factor of $(q^2)^j$.
The remaining $(2k - 2j)\, q_{\mu_i}$ factors will contract with the
$(2k - 2j)\, p^{\mu_i}$ factors to give $(p \cdot q)^{2k - 2j}$.
Since there are $(2k)!/[2^j j! (2k - 2j)!]$ terms in
$\{g \cdots g \hspace{1 pt} p \cdots p \}_{k, j}$, we find that
\begin{eqnarray}
&&\sum_{k = 1}^\infty
   \left( -g^{\mu \nu} q_{\mu_1} q_{\mu_2} C^{2k}_1 \right)
   q_{\mu_3} \cdots q_{\mu_{2k}} \frac{2^{2k}}{Q^{4k}}\,
   A_{2k} \Pi^{\mu_1 \cdots \mu_{2k}}			\nonumber\\
&&= -g^{\mu \nu} \sum_{k = 1}^\infty \sum_{j = 0}^k (-1)^j 
   \frac{(2k - j)!}{2^j (2k)!} \frac{(2k)!}{2^j j! (2k - 2j)!}
   (p^2\, q^2)^j\, (p \cdot q)^{2k - 2j}\,
  \frac{2^{2k}}{(Q^2)^{2k}}\, C^{2k}_1\, A_{2k}.
\end{eqnarray}

The $g^\mu_{\mu_1} g^{\nu}_{\mu_2} Q^2 C^{2k}_2$ term of
Eq.~(\ref{eq:T}) contributes to the coefficient of $-g^{\mu \nu}$
due to the identity
$g^\mu_{\mu_1} g^{\nu}_{\mu_2} g^{\mu_1 \mu_2} = g^{\mu \nu}$.
For fixed $k \in \mathbb{N}$ and $j \in \{0, \cdots, k\}$,
we seek the terms of
$\{g \cdots g \hspace{1 pt} p \cdots p \}_{k, j}$ which include
a factor of $g^{\mu_1 \mu_2}$.
The number of such terms is determined by the number of ways to
distribute the indices $\mu_3, \cdots, \mu_{2k}$ among $(j - 1)\, g$'s
and $(2k - 2j)\, p$'s without creating duplicate products.
{}From the $(2k)!/[2^j j (2k - 2j)!]$ ways to distribute the indices
$\mu_1, \cdots, \mu_{2k}$ over $j\, g$'s and $(2k - 2j)\, p$'s without
creating duplicates, relabeling indices
	$j \to j - 1$ and $k \to k - 1$
gives $(2k - 2)!/[2^{j - 1} (j - 1)! (2k - 2j)!]$ terms of
$\{g \cdots g \hspace{1 pt} p \cdots p \}_{k, j}$ which contain
$g^{\mu_1 \mu_2}$ for $k \in \mathbb{N}, j \in \{1, \cdots, k \}$.
(Note that there are no terms of
$\{g \cdots g \hspace{1 pt} p \cdots p \}_{k, 0}$ that contain
$g^{\mu_1 \mu_2}$.)
As for the $C^{2k}_1$ terms, we then find
\begin{eqnarray}
&&\sum_{k = 1}^\infty
   \left(g^\mu_{\mu_1} g^{\nu}_{\mu_2} Q^2 C^{2k}_2 \right)
   q_{\mu_3} \cdots q_{\mu_{2k}} \frac{2^{2k}}{Q^{4k}}\,
   A_{2k}\, \Pi^{\mu_1 \cdots \mu_{2k}}			\nonumber\\
&&= g^{\mu \nu} \sum_{k = 1}^\infty \sum_{j = 1}^k (-1)^j 
   \frac{(2k - j)!}{2^j (2k)!}
   \frac{(2k - 2)!}{2^{j - 1} (j - 1)! (2k - 2j)!}\,
   (p^2)^j (q^2)^{j - 1}\,
   (p \cdot q)^{2k - 2j} \frac{2^{2k}}{(Q^2)^{2k}}\, Q^2\,
   C^{2k}_2\, A_{2k}					\nonumber\\
&&\hspace{3.5 mm} +\ \text{terms not involving } g^{\mu \nu}.
\end{eqnarray}
Now, the term
$- i \epsilon^{\mu \nu \alpha \beta} g_{\alpha \mu_1} q_{\beta} q_{\mu_2}
   C^{2k}_3
 + (q^\mu q^\nu/Q^2) q_{\mu_1} q_{\mu_2} C^{2k}_4
 + \left( g^\mu_{\mu_1} q^\nu q_{\mu_2} + g^\nu_{\mu_1} q^\mu q_{\mu_2}
   \right) C^{2k}_5$
in Eq.~(\ref{eq:T}) will not contribute to $T_1$, as the indices
$\mu$ and $\nu$ are ``locked up'' in such a way that they cannot
result in a factor of $g^{\mu \nu}$ through contractions.
We conclude, therefore, that the coefficient $T_1$ of $-g^{\mu \nu}$
in the expansion (\ref{eq:T}) is
\begin{equation} \label{eq:T1}
\begin{split}
T_1
&= \sum_{k = 1}^\infty \sum_{j = 0}^k (-1)^j
   \frac{(2k - j)!}{2^j (2k)!} \frac{(2k)!}{2^j j! (2k - 2j)!} 
   (p^2 q^2)^j (p \cdot q)^{2k - 2j} \frac{2^{2k}}{(Q^2)^{2k}}\,
   C^{2k}_1\, A_{2k}						\\
&- \sum_{k = 1}^\infty \sum_{j = 1}^k (-1)^j
   \frac{(2k - j)!}{2^j (2k)!} \frac{(2k - 2)!}
   {2^{j - 1} (j - 1)! (2k - 2j)!}\,
   (p^2)^j (q^2)^{j - 1} (p \cdot q)^{2k - 2j}
   \frac{2^{2k}}{(Q^2)^{2k}} Q^2\, C^{2k}_2\, A_{2k}.
	\end{split}
\end{equation}
Substituting $p^2 = M^2$, $q^2 = -Q^2$, and $p \cdot q = Q^2/2x$ into
Eq.~(\ref{eq:T1}), changing indices in each term to $l = k - j$ and $j$,
and rearranging, gives the result
\begin{eqnarray}
T_1(x,Q^2)
&=& \sum_{l = 0}^\infty \sum_{j = 0}^\infty
   \binom{2l + j}{j}\,
   \mu^j\,
   \frac{1}{x^{2l}}\,
   C^{2l + 2j}_1\, A_{2l + 2j}			\nonumber\\
& &
 +\ \sum_{l = 0}^\infty \sum_{j = 1}^\infty
   \binom{2l + j}{j}
   \frac{j}{(l + j)(2l + 2j - 1)}\,
   \mu^j\,
   \frac{1}{x^{2l}}\,
   C^{2l + 2j}_2\, A_{2l + 2j}.
\end{eqnarray}
Finally, using the identity
$\oint_C d\omega\, \omega^{n - m - 1} = 2\pi i\, \delta_{nm}$
together with Eq.~(\ref{eq:disc}), we find that
\begin{eqnarray}
M^{(n)}_1(Q^2)
&=& \frac{1}{2} \frac{1}{2\pi i} \oint_C d\omega\,
    \frac{T_1 \left(1/\omega,Q^2\right)}{\omega^{n + 1}}
							\nonumber\\
% &=& \frac{1}{2} \sum_{l = 0}^\infty \sum_{j = 0}^\infty
%       \binom{2l + j}{j} \left( \frac{M^2}{Q^2} \right)^j
%       C^{2l + 2j}_1 A_{2l + 2j}
%       \frac{1}{2\pi i} \oint_C \omega^{2l - n - 1} \, d\omega \\
% & \hspace{4 mm} + \frac{1}{2} \sum_{l = 0}^\infty \sum_{j = 1}^\infty
%   \binom{2l + j}{j} \frac{j}{(l + j)(2l + 2j - 1)}
%   \left(\frac{M^2}{Q^2}\right)^j
%   C^{2l + 2j}_2  A_{2l + 2j} \frac{1}{2\pi i}
% \oint_C \omega^{2l - n - 1} \, d\omega
%
&=& \frac{1}{2} \sum_{l = 0}^\infty \sum_{j = 0}^\infty\,
    \mu^j\,
    \binom{2l + j}{j}\,
    C^{2l + 2j}_1\, A_{2l + 2j}\, \delta_{2l, n}		\nonumber\\
& &
 +\ \frac{1}{2} \sum_{l = 0}^\infty \sum_{j = 1}^\infty\,
    \mu^j\,
    \binom{2l + j}{j} \frac{j}{(l + j)(2l + 2j - 1)}\,
    C^{2l + 2j}_2\, A_{2l + 2j}\, \delta_{2l, n}
%
% &=& \frac{1}{2} \sum_{j = 0}^\infty \binom{n + j}{j}
%    \left( \frac{M^2}{Q^2} \right)^j 
%  C^{n + 2j}_1 A_{n + 2j} +
%  \frac{1}{2} \sum_{j = 1}^\infty \binom{n + j}{j}
%  \frac{2j}{(n + 2j)(n + 2j - 1)} 
%  \left(\frac{M^2}{Q^2}\right)^j C^{n + 2j}_2  A_{n + 2j}.
%
\label{eq:M1app}
\end{eqnarray}
which leads to Eq.~(\ref{eq:M1}).
The results for the other moments (\ref{eq:Mi_def}) are derived
in a similar manner.

%%%%%%%%%%%%%%%%%%%%%%%%%%%%%%%%%%%%%%%%%%%%%%%%%%%%%%%%%%%%%%%%%%%%%%%%%
\section{Structure function inversion}
\label{app:sf}

In this section we illustrate the standard moment inversion procedure
by presenting a detailed derivation for the case of the $F_1$ structure
function.  The derivations for the other structure functions can be
deduced straightforwardly from this example.

We begin by denoting the series in Eq.~(\ref{eq:M1}) involving
$C_1^{n+2j}$ and $C_2^{n+2j}$ by $m_1^{(n)}(Q^2)$ and $m_2^{(n)}(Q^2)$,
respectively.  Using Eq.~(\ref{eq:f_i}), we find for the $m_1^{(1)}$
term,
\begin{eqnarray}
\hspace*{-0.3cm}
\frac{1}{2\pi i} \int_{-i\infty}^{i\infty} dn\, x^{-n}\, m_1^{(n)}(Q^2)
&=& \frac{1}{4\pi i}
    \int_{-i\infty}^{i\infty} dn\,
    \int_0^1 dy\, x^{-n}\, y^{n-1}\, f_1(y)
    \sum_{j=0}^\infty \binom{n+j}{j}\,
    (\mu y^2)^j.
\end{eqnarray}
{}From the (generalized) binomial theorem, it follows then that
\begin{eqnarray}
\frac{1}{2\pi i} \int_{-i\infty}^{i\infty} dn\, x^{-n}\, m_1^{(n)}(Q^2)
&=& \frac{1}{4\pi i}
    \int_{-i \infty}^{i \infty} dn\,
    \int_0^1 dy\, x^{-n}\, y^{n-1}\, f_1(y)\,
    \frac{1}{(1 - \mu y^2)^{n + 1}}			\nonumber\\
&=& \frac{1}{2}
    \int_0^1 dy\, \frac{f_1(y)}{y(1 - \mu y^2)}\,
    \delta\left( \ln \frac{y}{x(1 - \mu y^2)} \right),
\end{eqnarray}
where we have used the $\delta$-function representation
\begin{equation}
\delta(\ln u) 
= \frac{1}{2\pi} \int_{-\infty}^{\infty} dn\, e^{in (\ln{u})}
= \frac{1}{2\pi i} \int_{-i\infty}^{i\infty} dn\, u^n.
\label{eq:delta-fn}
\end{equation}
Using the relation
\begin{equation}
\delta(u(y)) = \sum_{a \text{ = root of } u}
	       \frac{1}{|u'(a)|} \delta(y - a)
\end{equation}
with
$u(y)  = \ln(y/[x(1 - \mu y^2)])$ and
$u'(y) = (1 + \mu y^2)/[y(1 - \mu y^2)]$,
the only root of $u$ on $[0, 1]$ corresponds to $\xi = 2x/(1 + \rho)$,
which leads to
\begin{align}
\frac{1}{2\pi i} \int_{-i\infty}^{i\infty} dn\, x^{-n}\, m_1^{(n)}(Q^2)
&= {1 \over 2} \int_0^1 dy\,
   {f_1(y) \over 1 + \mu y^2}\,
   \delta(y - \xi).
\label{eq:m1app}
\end{align}

For the $m_2^{(n)}$ term, using integration by parts with
Eqs.~(\ref{eq:gi}) and (\ref{eq:parts}), we can write
\begin{equation}
\frac{1}{2\pi i} \int_{-i\infty}^{i\infty} dn\, x^{-n}\, m_2^{(n)}(Q^2)
= \frac{1}{2\pi i}
  \int_{-i \infty}^{i \infty} dn\,
  \int_0^1 dy\, x^{-n}\, y^{n - 2}\, g_2(y)
  \sum_{j = 1}^\infty j \binom{n + j}{j}\,
  (\mu y^2)^j.
\end{equation}
Next, from the relations
\begin{eqnarray}
\sum_{j = 1}^\infty j \binom{n + j}{j}\, (\mu y^2)^j
&=& \sum_{j = 1}^\infty\,
    (n + 1) \binom{n + j}{j - 1}\, (\mu y^2)^j		\nonumber\\
&=& (n + 1) {\mu\, y^2 \over (1 - \mu y^2)^{n+2}}, 
\end{eqnarray}
we obtain
\begin{equation}
{1 \over 2\pi i} \int_{-i\infty}^{i\infty} dn\, x^{-n}\, m_2^{(n)}(Q^2)
= {1 \over 2\pi i}\, \mu
  \int_{-i \infty}^{i \infty} dn\,
  \int_0^1 dy\, (n + 1)\, x^{-n} y^n\,
  {g_2(y) \over (1 - \mu y^2)^{n+2}}.
\end{equation}
Finally, since 
$(n + 1) x^{-n} = -x^2 (\partial/\partial x)\, x^{-n - 1}$,
we arrive at the result for the $m_2^{(n)}$ moment,
\begin{eqnarray}
\hspace*{-0.3cm}
\frac{1}{2\pi i} \int_{-i\infty}^{i\infty} dn\, x^{-n}\, m_2^{(n)}(Q^2)
&=& - \frac{1}{2\pi i}\, \mu x^2
    \frac{\partial}{\partial x}
    \int_{-i \infty}^{i \infty} dn\,
    \int_0^1 dy\,
    \frac{x^{-n - 1}\, y^n\, g_2(y)}{(1 - \mu y^2)^{n+2}}
							\nonumber\\
& & \hspace*{-0.5cm}
 =\ -\mu x^2
    \frac{\partial}{\partial x}
    \int_0^1 dy\,
    \frac{g_2(y)}{x(1 - \mu y^2)^2}
    \left[ \frac{1}{2\pi i} \int_{-i \infty}^{i \infty} dn\,
	   \left( \frac{y}{x(1 - \mu y^2)} \right)^n
    \right].						\nonumber\\
& &
% &=& -\mu x^2
%     \frac{\partial}{\partial x}
%     \left[ \frac{g_2(\xi)}{1 + M^2 \xi^2/Q^2} \right].
%
\label{eq:m2app}
\end{eqnarray}
Combining Eqs.~(\ref{eq:m1app}) and (\ref{eq:m2app}) then gives
the final result for the inverted $F_1$ structure function,
\begin{equation}
F_1(x,Q^2)
= \frac{1}{2(1 + \mu \xi^2)} f_1(\xi)
- \mu x^2
  \frac{\partial}{\partial x} 
  \left( \frac{g_2(\xi)}{1 + \mu \xi^2} \right).
\end{equation}
The results for the other structure functions $F_2, \cdots, F_5$
in Eqs.~(\ref{eq:Fi_full}) follow analogous derivations.

%%%%%%%%%%%%%%%%%%%%%%%%%%%%%%%%%%%%%%%%%%%%%%%%%%%%%%%%%%%%%%%%%%%%%%%%%

\end{document}